\documentstyle[11pt,amstex]{article}                                

\def\AFOUR{%
\setlength{\textheight}{9.0in}%
\setlength{\textwidth}{5.75in}%
\setlength{\topmargin}{-0.375in}%
\hoffset=-.5in%
\renewcommand{\baselinestretch}{1.17}%
\setlength{\parskip}{6pt plus 2pt}%
}
\AFOUR                                           
\makeatletter
\def\section{\@startsection {section}{1}{\z@}{-3.5ex plus -1ex minus
 -.2ex}{2.3ex plus .2ex}{\large\bf}}
\def\subsection{\@startsection{subsection}{2}{\z@}{-3.25ex plus -1ex minus
 -.2ex}{1.5ex plus .2ex}{\normalsize\bf}}
\makeatother
\makeatletter
\@addtoreset{equation}{section}

\makeatother
\newsymbol\ltimes226E
\newcommand{\nc}{\newcommand}
\newcommand{\rnc}{\renewcommand}
\nc{\bea}{\begin{eqnarray}}
\nc{\eea}{\end{eqnarray}}
\nc{\be}{\bea}
\nc{\ee}{\eea}

\def\slash#1{\setbox0=\hbox{$#1$}#1\hskip-\wd0\hbox to\wd0{\hss\sl/\/\hss}}

\def\href#1#2{{#2}}

\rnc{\a}{\alpha}
\nc{\ab}{\bar{\a}}
\nc{\ap}{\a^{+}}
\nc{\abm}{\ab^{-}}
\rnc{\b}{\beta}
\nc{\bb}{\bar{\b}}
\nc{\bbp}{\bb_{\zb}^{+}}
\nc{\bm}{\b_{z}^{-}}
\nc{\oa}{\overline{\a}}
\nc{\ob}{\overline{\b}}
\rnc{\gg}{\gamma}
\rnc{\d}{\delta}
\nc{\f}{\phi}
\nc{\fb}{\bar{\phi}}
\nc{\vf}{\varphi}
\nc{\p}{\psi}

\rnc{\c}{\chi}
\nc{\la}{\lambda}
\nc{\m}{{\mathrm m}}
\nc{\n}{\nu}
\rnc{\o}{\omega}
\nc{\Om}{\Omega}
\rnc{\t}{\theta}
\nc{\eps}{\epsilon}
\rnc{\S}{\Sigma}
\nc{\F}{\Phi}
\nc{\trac}[2]{{\textstyle\frac{#1}{#2}}}
\nc{\ex}[1]{\mbox{e}^{\,\textstyle#1}}
\nc{\mat}[4]{\left(\begin{array}{cc}#1&#2\\#3&#4\end{array}\right)}
\nc{\som}[9]{\left(\begin{array}{ccc}#1&#2&#3\\#4&#5&#6\\#7&#8&#9%
\end{array}\right)}
\nc{\tr}{\mathop{\mbox{tr}}\nolimits}
\nc{\ad}{\mathop{\mbox{ad}}\nolimits}
\nc{\Tr}{\mathop{\mbox{Tr}}\nolimits}
\nc{\Det}{\mathop{\mbox{Det}}\nolimits}
\nc{\rk}{\mathop{\mbox{rk}}\nolimits}
\nc{\ra}{\rightarrow}
\nc{\Ra}{\Rightarrow}
\nc{\LRa}{\Leftrightarrow}
\nc{\ot}{\otimes}
\rnc{\ss}{\subset}
\nc{\nul}{\noindent\underline}
\nc{\non}{\nonumber\\}
\nc{\subs}[1]{{\vspace*{0.5cm}}%
{\noindent\underline{#1}}{\addcontentsline{toc}{subsection}{#1}}%
{\vspace*{0.3cm}}}
\nc{\zb}{\bar{z}}
\rnc{\lg}{\frak{g}}
\nc{\lt}{\frak{t}}
\nc{\lk}{\frak{k}}
\nc{\lh}{\frak{h}}
\nc{\pik}{\Pi_{\lk}}
\nc{\pip}{\Pi_{+}}
\nc{\pim}{\Pi_{-}}
\nc{\pih}{\Pi_{\lh}}
\nc{\jz}{J_{z}}
\nc{\jzh}{\jz^{\lh}}
\nc{\jzp}{\jz^{+}}
\nc{\jzm}{\jz^{-}}
\nc{\del}{\partial}
\nc{\dz}{\del_{z}}
\nc{\dzb}{\del_{\bar{z}}}
\nc{\az}{A_{z}}
\nc{\azb}{A_{\bar{z}}}
\nc{\g}{g^{-1}}
\nc{\dw}{\Delta_{W}}
\nc{\Ad}{{\mbox{Ad}}}
\nc{\ks}{Ka\-za\-ma-\-Su\-zu\-ki}
\nc{\KS}{\ks}
\nc{\ksm}{\ks\ model}
\rnc{\AA}{{\Bbb A}}
\nc{\BB}{{\Bbb B}}
\nc{\CC}{{\Bbb C}}
\nc{\PP}{{\Bbb P}}
\nc{\cpm}{\CC\PP(m)}
\nc{\cpn}{\CC\PP(n)}
\nc{\cp}[1]{\CC\PP(#1)}
\nc{\gmn}{G(m,m+n)}
\nc{\gmnk}{\gmn_{k}}
\nc{\cO}{{\cal O}}
\nc{\bcO}{\bar{\cO}}
\nc{\bO}{\bar{O}}
\nc{\oQ}{\overline{Q}}
\begin{document}
\global\parskip=4pt
\makeatother\begin{titlepage}
\begin{flushright}
\rightline{ICTP/98/1, hep-th/98}
\end{flushright}
\vspace*{0.5in}
\begin{center}
{\LARGE\sc  Bound States Of Type I D-Strings}\\
\vskip .3in
\makeatletter
\begin{tabular}{c}
{\bf E. Gava}, \footnotemark 
\\ 
INFN and ICTP, Trieste, Italy, \\
\end{tabular}

\begin{tabular}{c}
{\bf J.F. Morales}, \footnotemark
\\ 
SISSA, Trieste, Italy,  \\
\end{tabular}

\begin{tabular}{c}
{\bf K.S. Narain} and {\bf G. Thompson}\footnotemark
\\ ICTP, P.O. Box 586, 34014 Trieste, Italy\\
\end{tabular}
\end{center}
\addtocounter{footnote}{-2}%
\footnotetext{e-mail: gava@@isolde.sissa.it}
\addtocounter{footnote}{1}%
\footnotetext{e-mail: morales@@sissa.it}
\addtocounter{footnote}{1}%
\footnotetext{e-mail: narain,thompson@@ictp.trieste.it}
\addtocounter{footnote}{1}%
\vskip .50in
\begin{abstract}
\noindent 
We study the infra-red limit of the $O(N)$ gauge theory that describes the
low energy modes of a system of $N$ type I D-strings and provide some
support to the conjecture that, in this limit, the theory flows to an
orbifold conformal theory. We compute the elliptic genus of the orbifold
theory and argue that its longest string sector describes the bound states
of D-strings. We show that, as a result, the masses and multiplicities of
the bound states are in agreement with the predictions of heterotic-type I
duality in 9 dimensions, for all the BPS charges in the lattice
$\Gamma_{(1,17)}$. 

\end{abstract}
\makeatother
\end{titlepage}
\begin{small}
\end{small}

\setcounter{footnote}{0}

\section{Introduction}

The putative duality between Type I string theory and the $SO(32)$
heterotic string theory \cite{Witt,pw} requires the existance of Type I
D-string bound states. The way to see this is to begin by compactifying on
a circle of radius $R_H$ in the $X^{9}$ direction. On the heterotic side
the electric charge spectrum in the nine dimensional theory sits in the
lattice $\Gamma_{(1,17)}$. This lattice arises on taking into account the
charges associated with the gauge fields of the Cartan subalgebra of
$SO(32)$, the $G_{\mu 9}$ component of the metric and the $B_{\mu 9}$
component of the Neveu-Schwarz antisymmetric tensor. In particular, the
states carrying $N$ units of $B_{\mu 9}$ charge correspond to the
fundamental heterotic string wrapping $N$ times around the $X^{9}$ circle.
This particular charge corresponds to a null vector in $\Gamma_{(1,17)}$
and scrutiny of the partition function shows that the multiplicity of such
BPS states, which arise at the level one oscillator mode of the bosonic
sector, is $24$. The general BPS spectrum, with all the electric charges,
is given by
\be
kN - \frac{1}{2}P^{2}=N_{R}-1 ,
\label{level}
\ee
where $P$ is a vector in $\Gamma_{(0,16)}$, $N_{R}$ is the bosonic
oscillator number and $k$ is an integer, related to the Kaluza-Klein
momentum $p_9$ carried by the state in the following way:  
\be 
p_9 = \frac{1}{R_H}(k + B.P + \frac{1}{2}B^{2}N), 
\label{kkmom} 
\ee 
where $B^{i}$ are the holonomies in the Cartan subalgebra of $SO(32)$ in
the $X^{9}$ direction (Wilson lines).  The multiplicity is then given by
the $N_{R}$'th oscillator level in the partition function $\eta^{-24}$. 
Furthermore, the mass $m$ in the string frame is given by
\be 
m=\left| p_9 + \frac{NR_H}{\alpha'} \right| .  
\label{mass} 
\ee

On the other hand, the Type I theory is related to the heterotic by a
strong-weak duality. The coupling constants and metrics are related by
\be
\la_{I} = \frac{1}{\la_{H}} , \;\;\; G^{I}_{MN} =\frac{G^{H}_{MN}}{
\la_{H}} .
\label{drel}
\ee
The duality relations imply that a heterotic state labelled by $N$, $k$
and $P$ is mapped to a type I state with Kaluza-Klein momentum $p_9$ and
mass $m$, in the string frame, given by: 
\begin{eqnarray}
p_9 &=& \frac{1}{R_I}(k + B.P + \frac{1}{2}B^{2}N)
\label{kkmomI} \\
m &=& \left| p_9 + \frac{NR_I}{\alpha'\lambda_I} \right| ,
\label{massI} 
\end{eqnarray}
where $R_I$ is the type I radius along the $X^9$ direction. 

The Neveu-Schwarz two-form, $B_{MN}$, of the heterotic string is mapped to
the Ramond-Ramond antisymmetric tensor field, $B_{MN}^{R}$, of the Type I
string. Consequently, the winding modes of the fundamental string on the
heterotic side are mapped to the D-string winding modes on the Type I
side. Duality, therefore, predicts that a BPS system of $N$ D-strings, in
the Type I theory, each of which is wrapped once around the circle, should
have a threshold bound state with multiplicity equal to $24$. Furthermore,
when there are other charges turned on, the multiplicities of the
resulting $D$-string threshold bound states should reproduce the heterotic
multiplicity (\ref{level}).

Our aim in this letter is to establish that this is indeed the case. We
begin by reviewing the effective world volume gauge theory of $N$ type I
D-strings and argue that in the infrared limit, this gauge theory flows to
an orbifold conformal field theory \cite{BM,Lowe,Rey}, in analogy with a
similar phenomenon in the type II case \cite{HMS,BJSV,DVV,M,BS}. We
compute the elliptic genus of this conformal field theory and show that
the twisted sector corresponding to the longest string reproduces the
results expected from heterotic-type I duality.  We also argue that the
threshold bound states arise precisely in the longest string sector. 

\section{Type-I D-Strings}
We begin by recalling the form of the world volume action which describes
the low lying modes of a system of $N$ D-strings in the type I
theory: 
\bea
S& =& \Tr \int d^{2}x \; -\frac{1}{4g^{2}}F^{2} + (DX_{I} )^{2} +
g^2([X_{I},X_{J}])^{2} \nonumber \\
& & \;\;\; + \Lambda
\slash{D} \Lambda + S \slash{D} S + \sum_{a=1}^{16} \bar{\chi}^{a} \slash{D}
\chi^{a} + g\Lambda \Gamma^I [X_I, S]  + \sum_{a=1}^{16} \bar{\chi}^{a} B^a
\chi^a.
\label{Iaction}
\eea 
The fields transform in various representations of the gauge group
$O(N)$. $X$ and $S$ transform as second rank symmetric tensors, while
$\Lambda$ and $\chi$ transform in the adjoint and fundamental
representations respectively. There is an $SO(8)_{{\cal R}}$, R
symmetry group, under which $X$, $S$, $\Lambda$ and $\chi$ transform
as an ${\mathbf 8}_V$ (this is the $I$ label), an ${\mathbf 8}_S$,
an ${\mathbf 8}_c$ and a singlet, respectively. The
$\chi$ transforms under the $SO(32)$ in
the vector representation with $\chi^a$ and $\bar{\chi}^a$ denoting the
positive and negative weights. The $\Lambda$ and
$\chi$ are negative chiral (right-moving) world sheet fermions while
the $S$ are positive chiral (left-moving) fermions. Finally $B^a$ are the 
background holonomies (i.e. Wilson lines on the 9-branes) in the Cartan
subalgebra of $SO(32)$. The Yang-Mills coupling $g$ is related to the type
I string coupling  via $g^2 = \lambda_I/\alpha'$. The
vev's of the $X$ fields,
appearing in the action above, measure the distances between the D-strings
in units of $\sqrt{\alpha' \lambda_I}$. This fact will be important later,
when we compare the spectrum with that of the heterotic theory.

Geometrically the fields appear in the following fashion \cite{pw,DF}. 
The above action arises as the $Z_{2}$ projection of the corresponding
theory in the type II case. Recall that in the type II situation a system
of $N$ branes has a $U(N)$ symmetry. Write the hermitian matrices as a sum
of real symmetric matrices and imaginary anti-symmetric matrices. The
$Z_{2}$ projection, for type I $D$-strings, assigns to the world volume
components of the gauge field the anti-symmetric matrices, that is, it
projects out the real symmetric part and so reduces the gauge group to
$O(N)$. On the other hand, the components of the gauge field in the
transverse directions, $X$, have their imaginary part projected out and so
are symmetric matrices transforming as second rank symmetric tensors under
the $O(N)$. The diagonal components of the $X$ give the positions for the
$N$ branes. The trace part, which we factor out, represents the center of
mass motion.

The $\chi$ carry the $SO(32)$ vector label, as they are the lowest modes
of the strings which are stretched between the $9$-branes and the
$D$-strings. 

As mentioned earlier, the winding mode $N$ of the fundamental heterotic
string is mapped, via duality, to $N$ D-strings on the Type I side that
wind around the $X^9$ circle. Since the former appears as a fundamental
BPS state with multiplicity $24$, we should find that the system of $N$
D-strings in type I theory admits threshold bound states with multiplicity
$24$. In other words, in the $O(N)$ theory, there should be $24$ square
integrable ground states that are 10 dimensional $N=1$ vector short BPS
super-multiplets. That every ground state appears with 8 bosonic and 8
fermionic modes, necessary to form the $N=1$ short vector supermultiplet,
follows from the the fact that there are zero modes of the $O(N)$ singlet
free field $S$, describing the center of mass motion (which has not been
included in the action (\ref{Iaction})). The remaining part of the $O(N)$
theory described in the action (\ref{Iaction}), therefore, must have, $24$
bosonic normalizable ground states as predicted by the heterotic-type I
duality. In other words, we want to show that the Witten Index for the
above theory is $24$.

More generally, in the heterotic theory, we can also turn on other
charges, namely the one associated with Kaluza-Klein modes that couple to
$G_{\mu 9}$ and the ones associated with the Cartan subalgebra of the
$SO(32)$ gauge group. These charges can also be excited in the system of
$N$ D-strings in the type I theory.  Indeed, one can include states
carrying a longitudinal momentum along the string and thereby generate
Kaluza-Klein momentum. Similarly, one can generate $SO(32)$ quantum
numbers by suitably applying $\chi$ modes. The information about the
multiplicities of states carrying these extra charges will be contained in
the elliptic genus of the above theory (\ref{Iaction}). 

Since the Witten index and, more generally, the elliptic genus do not
depend on the coupling constant, we can take a limit which is most
convenient for our present purposes. We will consider the infra-red limit
of the theory, as it has been conjectured in \cite{BM,Lowe,Rey}, that in
this limit the theory flows to a $(8,0)$ orbifold superconformal field
theory. This is in analogy with a similar conjecture for a system of type
IIB D-strings \cite {DVV}. In the following we give some support to this
conjecture by, first, gauge fixing (\ref{Iaction}) and then by performing
a formal scaling which yields the orbifold theory directly.

\section{Type II and the IR Limit}

Before discussing the type I theory we make a digression on the type II
theory that will prove useful later. Our aim here is to show that with a
prudent choice of gauge one can simplify matters considerably. This
prepares the way for taking the large coupling limit in a fashion that is,
to a large extent, controlable. We will work with a little more generality
than is really required and begin with an analysis of $D$ dimensional
Yang-Mills theory reduced to $d$ dimensions. $D$ dimensional vector labels
are denoted by $M,N, \dots$, those in $d$ dimensions are denoted by $\mu,
\nu, \dots$ and those in the remaining $D-d$ (reduced) dimensions by $I,J,
\dots$. To make contact with the type II $D$-brane world volume theories
one sets $D=10$. 

The $D$ dimensional Yang-Mills theory has a `potential' of the form,
\be
-\tr \, \frac{g^{2}}{4}[A_{M},A_{N}]^{2} .
\ee
Minimising, in any dimension $d$, we learn that we are interested in
the fields that live in the Cartan subalgebra. Decompose the Lie-
algebra, $\lg$, of the gauge group as $\lg = \lt \oplus \lk$, where
$\lt$ is the prefered Cartan subalgebra and $\lk$ is its ortho-
complement. It makes good sense, therefore, to perform a non-canonical
split,
\be
A_{M} = A_{M}^{\lt} + A_{M}^{\lk} ,
\ee
where the superscripts indicate the part of the Lie-algebra that the
fields live in. Before proceeding we need to gauge fix. Given the
splitting of the algebra, it behoves us to choose the `background field'
gauge\footnote{At this point the connection is $gA$, which explains some
of the, what appear to be, spurious factors of $g$. We are gauge fixing
the `canonical' gauge field and not the one scaled by $g$ so that the BRST
transformations are $Q(gA_{M}) = D_{M}(gA)C$ and $QC = C^{2}$.}
\be
D^{M}(gA^{\lt})gA^{\lk}_{M} =0 ,
\ee
which preserves the maximal Torus gauge invariance. The ghosts come in as
\be
\tr \overline{C}^{\lk}D_{M}(gA^{\lt})D^{M}(gA)C^{\lk} + \, \tr \,
g^{2}\overline{C}^{\lk} [ \, [A_{M}^{\lk}, C^{\lk}]^{\lt} , 
A^{M \, \lk} ] .
\ee
We choose a Feynman type gauge with a co-efficient chosen to give the
most straightforward analysis namely we add
\be
\tr -\frac{1}{2}\left( D^{M}(gA^{\lt})A^{\lk}_{M}\right)^{2}
\ee
to the action. With this choice the potential becomes
\be
-\tr \, \frac{g^{2}}{2}[A_{M}^{\lt},A_{N}^{\lk}]^{2} + \dots  ,
\ee
where the ellipses indicate higher order terms in $A_{M}^{\lk}$ and
which, directly, will be seen to be irrelevant.

We now perform the following sequence of scalings on the fields
appearing in a $N=1$ super Yang-Mills theory in $D$ dimensions
\be
A_{M}^{\lk} \rightarrow \frac{1}{g}A_{M}^{\lk} , \;\;\;\; \psi^{\lk}
\rightarrow \frac{1}{\sqrt{g}} \psi^{\lk}, \;\;\;\; \overline{C}^{\lk}
\rightarrow \frac{1}{g^{2}} \overline{C}^{\lk} . 
\ee
On a torus, $T^{d}$, with periodic boundary conditions on all the fields
appearing, this scaling has unit Jacobian. We can now take the $g
\rightarrow \infty$ limit. The action, in this limit is:
\bea
S &=& \tr \, \int d^{d}x -\frac{1}{4}F_{MN}(A^{\lt})^{2} + \psi^{\lt}
\slash{\partial} \psi^{\lt} 
-\frac{1}{2}[A_{M}^{\lt},A_{N}^{\lk}]^{2}\nonumber\\
& &\;\;\;+ \psi^{\lk}
\Gamma^{M}[A_{M}^{\lt}, \psi^{\lk}] + [\overline{C}^{\lk},
A_{M}^{\lt}][C^{\lk}, A_{M}^{\lt}] .
\eea
All the fields in the $\lk$ part of the Lie-algebra can be integrated
out and clearly give an overall contribution of unity to the path
integral. Thus, we are left with a free, supersymmetric, system of Cartan
valued fields. By invoking the Weyl symmetry
that is left over, one finds that the target space of the theory is
$({\mathbf R}^{(D-d)r}\times T^{dr})/ W$, where $r$ is the rank of the
group and $W$ is the Weyl group.

Fixing $D=10$, gives us the Type II world volume theories of parallel
$D$ branes. The limit just described, the strong coupling limit in
the gauge theory, when $d=2$ and $D=10$ (the D-string) gives rise to 
the orbifold conformal field theory as in \cite{DVV}.

\section{Type I and the IR Limit}

The flat directions of the potential in this case require mutually
commuting matrices once more. We denote those $X$'s, with a slight abuse of
notation, by $X^{\lt}$ (for example one may choose these to be
diagonal). A convenient way to proceed is to start with the (complexified) 
$SU(N)$ Lie algebra and to split it into a Cartan subalgebra $\lt$ and 
into positive and negative roots, $\lk_{+}$ and $\lk_{-}$, respectively, 
that is, $\lk = \lk_{+} \oplus \lk_{-}$. The $Z_{2}$ projection
means that, in this basis, the world volume gauge fields are
proportional to the anti-symmetric (imaginary part of $\lk$) generators, 
$\m_{-}=\lk_{+}-\lk_{-}$, while the
$X$'s are proportional to the symmetric  
generators, $\lt$ and (real part of $\lk$) $\m_{+}=
\lk_{+} + \lk_{-}$. With
these identifications the bosonic parts of the type I and type II
theories coincide. We choose the same gauge fixing as in the type II
theory, now restricted to the $\m_{-}$ directions,
\be
g \partial^{\mu}A_{\mu}^{\m_{-}} + g^2 [X^{\lt}, X^{\m_{+}}] =0
\label{gfI} 
\ee
and we scale the fields in a similar way, that is
\bea
& & A_{\mu}^{\m_{-}} \rightarrow \frac{1}{g} A_{\mu}^{\m_{-}},
\;\;\;\; X^{\m_{+}}_{I} \rightarrow \frac{1}{g} X^{\m_{+}},
\;\;\;\; \Lambda^{\m_{-}} \rightarrow
\frac{1}{\sqrt{g}}\Lambda^{\m_{-}}, \nonumber \\
& &  \;\;\; \; \; \; \; S^{\m_{+}}
\rightarrow \frac{1}{\sqrt{g}}S^{\m_{+}}, \;\;\;\;\; \overline{C}^{\m_{-}} 
\rightarrow \frac{1}{g^{2}}\overline{C}^{\m_{-}} .
\eea
The remaining fields $X^{\lt}$, $S^{\lt}$, $C^{\m_{-}}$ 
and $\chi$ are unchanged. As before the Jacobian of these
scalings is unity if we take periodic boundary conditions for the
fermions $S$ and $\Lambda$. There is no such requirement on the
$\chi$. Consequently the $g \rightarrow \infty$ limit may be safely taken. 

The action now takes the form
\bea
S & =& \tr \, \int d^{d}x -\frac{1}{2}
\left|\partial_{\mu}X_{I}^{\lt}\right|^{2} + S^{\lt}
\slash{\partial} S^{\lt} -\frac{1}{2}[X_{I}^{\lt},A_{\mu}^{\m_{-}}]^{2}
-\frac{1}{2}[X_{I}^{\lt},X_{J}^{\m_{+}}]^{2} \nonumber \\
& & \; + \Lambda^{\m_{-}}
\Gamma^{I}[X_{I}^{\lt}, S^{\m_{+}}] + [\overline{C}^{\m_{-}},
X_{I}^{\lt}][C^{\m_{-}}, X_{I}^{\lt}]  + \sum_{a=1}^{16}\bar{\chi}^a \slash{D}
\chi^a +\sum_{a=1}^{16}\bar{\chi}^a B^a
\chi^a.
\eea
Formally, since the $\chi$ fields are chiral, only the right moving part
of the gauge field is coupled to it and one can perform the integral over
the left moving part of the gauge field which sets the right moving part
to zero. Hence, on integrating out the massive modes, one would be left
with a completely free theory of the massless modes $X^{\lt}$, $S^{\lt}$
and $\chi$. The determinant factors would then, at least formally, cancel
between the fields of various statistics. 

However, the above cancellation of the determinant factors is a bit quick.
If correct, it would imply that even if we had started with an anomalous
theory we would end up, in the limit, with a well defined superconformal
field theory. For example, this would seem to be the case if we simply
ignored the $\chi$ fields altogether.  The point is that each fermionic
determinant appearing is anomalous. These determinants, when defined in a
vector gauge invariant way, involve extra quadratic terms in the gauge
field. The presence of these would mean that the functional determinants
would not cancel, since the gauge field contribution would not be $\Det
(X^{\lt})^{2}$. Happily, the condition that the theory be anomaly free
means that the total sum of these extra pieces is zero and this is exactly
what is required to make our formal argument above work. 

On including the center of mass one gets $N$ of the $X$'s and $S$'s, each
transforming as a ${\mathbf 8}_{V}$ and ${\mathbf 8}_{S}$ of $SO(8)$
respectively and $N$ $\chi$'s each transforming as a fundamental of
$SO(32)$. The field content is like that of $N$ copies of the heterotic
string in the light-cone gauge with an effective inverse tension
\be
\alpha'_{\mathrm eff} = \alpha' \lambda_I.
\label{eff}
\ee

The condition (\ref{gfI})  does not completely fix the gauge, there are
still discrete transformations which leave the action invariant. There is
the permutation group $S_{N}$ which permutes the $N$ copies of
$(X,S,\chi)$ and which has the interpretation of permuting the $N$
$D$-strings. There are also $O(N)$ transformations which leave invariant
$X$ and $S$ but which act non-trivially on the $\chi$'s by reflection
giving rise to a $Z_{2}^{N}$. The full orbifold group is therefore the
semidirect product $S_{N} {\ltimes} Z_{2}^{N}$. 

\section{Orbifold Partition Function}

We are interested in calculating the elliptic genus of the orbifold
conformal theory. In this case the $S$ fermions have periodic boundary
conditions on the world sheet torus. The elliptic genus for our conformal
field theory will be zero due to the fact that the center of mass $S$ will
have zero modes in all the twisted sectors as it is orbifold group
invariant. However, the zero modes of the center of mass $S$ precisely
give rise to the 8 bosonic and 8 fermionic transverse degrees of freedom
that fill out the 10-dimensional $N=1$ vector supermultiplet corresponding
to a BPS short multiplet. Our goal here is to calculate the multiplicities
of these BPS states, that are clearly governed by the elliptic genus of
the remaining conformal field theory that describes the relative motions
of the D-strings. This means that we need to consider only those twisted
sectors that have at most the zero modes of the center of mass $S$. 

Let us briefly review how the orbifold elliptic genus is computed
\cite{DHVW,DMVV}.  Each twisted sector of the orbifold corresponds to a
conjugacy class of the orbifold group.  A general element of the group $G=
S_N \ltimes Z_2^N$ can be denoted by $(g,\omega)$ where $g\in S_N$ and
$\omega \in Z_2^N$. First let us identify the twisted sectors where the
$S$'s have no other zero modes besides the center of mass one. For this it
is sufficient to consider the action of the elements of $S_N$ since the
$Z_2^N$ part does not act on $S$ fields.  A general conjugacy class $[g]$
in $S_N$ is characterized by partitions $\{ N_n\}$ of $N$ satisfying $\sum
nN_n = N$ where $N_n$ denotes the multiplicity of the cyclic permutation
$(n)$ of $n$ elements in the decomposition of $g$ as
\be
[g]=(1)^{N_1}(2)^{N_2}\cdots(s)^{N_s}   .
\ee
In the $[g]$-twisted sector the fields satisfy the boundary condition:
$(X, S, \chi) (\sigma+2\pi R_I ) = g (X, S, \chi)(\sigma)$ where 
$\sigma$ is the coordinate along the string. 

In each twisted sector one must project by the centralizer subgroup 
$C_g$ of $g$, which takes the form:
\be
C_g = \prod_{n=1}^s S_{N_n}\ltimes Z_n^{N_n}   ,
\ee
where each factor $S_{N_n}$ permutes the $N_n$ cycles $(n)$, while each
$Z_n$ acts within one particular cycle $(n)$. In the path integral
formulation this projection involves summing over all the boundary
conditions along the world-sheet time direction $t$, twisted by elements
$h$ of $C_g$. We shall denote by $([g],h)$ the twisted sector with twist
$g$ along the $\sigma$ direction and twist $h$ along the $t$ direction. 

We will now show that if $[g]$ involves cycles of different lengths, say
$(n)^a$ and $(m)^b$ with $n \neq m$, then the corresponding twisted sector
does not contribute to the elliptic genus. To see this, we note that there
are now at least two sets of zero modes for $S$, which can be expressed,
by a suitable ordering of indices, as $(S_1+ S_2 +\cdots S_{na})$ and
$(S_{na+1}+\cdots S_{na+mb})$, where the two factors $(n)^a$ and $(m)^b$
act on the two sets of indices in the obvious way. These zero modes
survive the group projection because the centralizer of $[g]$ does not
contain any element that mixes these two sets of indices with each other,
thereby giving zero contribution to the elliptic genus. Thus we need only
to consider those sectors with $[g]=(L)^M$ where $N=LM$. 

The centralizer in the case where $[g]=(L)^M$ is $C_g= S_M \ltimes Z_L^M$.
From the boundary condition along $\sigma$ it is clear that there are $L$
combinations of $S$'s that are periodic in $\sigma$. By suitable ordering,
they can be expressed as $S^k = \sum_{i=Lk+1}^{L(k+1)}S_i$ for $k=0,\dots,
M-1$. These zero modes have to be projected by the elements $h$ in the
centralizer $C_g$. In particular, when $h$ is the generator of $Z_M
\subset S_M \subset C_g$, it acts on the zero modes $S^k$ by cyclic
permutation. It is clear, therefore, that only the center of mass
combination $\sum_{k=0}^{M-1} S^k$ is periodic along the $t$ direction.
Hence, this sector contributes to the elliptic genus. More generally any
$h=(e,f)\in C_g=S_M \ltimes Z_L^M$ will satisfy the above criteria
provided $e = (M)\in S_M$ and $f$ is some element of $Z_L^M$. The number
of such elements $h$ is $(M-1)! \times L^M$.

In particular when $N$ is prime the sectors that contribute to the
elliptic genus are: $(1,h)$ where $h \in Z_N$ and $([g],h)$ with $[g]=(N)$
and $h$ in the corresponding centralizer $Z_N$. In the following we shall
refer to these two types of sectors as the shortest and longest string
sectors, respectively.

The full orbifold group $G$ is specified by an element of $S_N$ (discussed
above)  together with an element of $Z_2^N$ that acts on the $\chi$'s. Let
us denote a general element by $(g,\epsilon)$ where $g\in S_N$ and
$\epsilon \in Z_2^N$. Now $S_N$ acts as an automorphism in $Z_2^N$ by
permuting the various $Z_2$ factors. We denote this action by
$g(\epsilon)$. Then the semi-direct product is defined in the usual way: 
$(g,\epsilon).(g',\epsilon') = (gg', \epsilon g(\epsilon'))$. Twisted
sectors will now be labelled by a conjugacy class in $G$. The relevant
sectors, for the elliptic genus computation, as discussed above, are the
conjugacy classes $[g]$ in $S_N$ of the form $[g] = (L)^M$ with $N=LM$.
One can easily verify that the various classes in $G$ are labelled by
$([g], \epsilon)$ with $\epsilon =\epsilon_1.\epsilon_2\dots \epsilon_M$
where each $\epsilon_i$ is in the quotient subgroup of $Z_2^L$ by its even
subgroup.  Combining this with the condition that we have found for $h$ we
may conclude that all the $\epsilon_i$'s must be equal (i.e.  all
$\epsilon_i$'s must be either even or odd element of $Z_2^L$) in order for
such $h$ to exist in the centralizer of $([g],\epsilon)$ in $G$. In this
case the centralizer is the group of elements of the form $(h,\alpha)$
where $h\in C_g$ and $\alpha \in Z_2^N$ satisfies $\epsilon h(\epsilon) =
\alpha g(\alpha)$. The number of independent such $\alpha$'s is $2^M$ and
therefore the order of the centralizer of $([g],\epsilon)$ is $M! L^M
2^M$.

We now proceed to compute the elliptic genus $\tr(-1)^F \ex{- T H + 2\pi i
R \tau_1 P_{\sigma}}$ where $H$ and $P_{\sigma}$ are the Hamiltonian and
the longitudinal momentum.  The computation for general $L$ and $M$ is
quite tedious, therefore we will only describe it for the longest string
sector (i.e. $M=1$ and $L=N$).  The centralizer $C_g=Z_N$ and consists of
elements of the form $h= g^s$ for $s=0,1,\dots,N-1$ and as a result their
action is obtained by modular transformations $\tau_1 \rightarrow
\tau_1+s$ from the $h=1$ sector. Thus we can restrict ourselves to $h=1$.
The eigenvalues of $[g]$ are $\omega^r$ for $r=0,1,\dots,N-1$ where
$\omega =e^{2i\pi/N}$. As a result the $N$ copies of $X$'s and $S$'s come
with fractional oscillator modes that are shifted by $r/N$ in units of
$1/R_I$. The left moving part of the non-zero mode partition function
cancels between $X$'s and $S$'s. The zero modes appear for the center of
mass (i.e. $r=0$):  the zero modes of $S$ give rise to the usual 8 bosonic
and 8 fermionic degrees of freedom filling out the BPS vector
supermultiplet, while the zero modes of the left and right moving $X$'s
give a factor $\tau_2^{-4}$. The right moving $X$'s, upon taking the
product over $r$, give $1/\eta(q^{\frac{1}{N}})^8$, upto a zero point
shift, where $q = exp(2\pi i \tau)$ with $\tau = \tau_1 + i \frac{T}{2\pi
R_I} \equiv \tau_1 + i\tau_2$.

To include the contribution of the $\chi$'s we must specify the group
elements in $Z_2^N$ as well. There are two possible $\epsilon$'s that come
with $[g]$:  the even and odd element of $Z_2^N$. First, let us consider
the situation when the Wilson lines $B_a$ are set to zero. By taking the
product of all the eigenvalues of the twist, these for odd $N$, give rise,
respectively, to the Neveu-Schwarz and Ramond sectors of the $SO(32)$
fermions, with $q$ replaced by $q^{\frac{1}{N}}$ (again upto zero point
shifts). For even $N$, on the other hand, the Neveu-Schwarz and Ramond
sectors appear for $\epsilon$ odd and even, respectively.  Furthermore the
centralizer contains two elements with $h=1$ namely $\alpha = \pm 1$.
These two choices give rise to the usual GSO projection. 

Finally, one can compute the zero point shift for the right moving $X$'s
and $\chi$'s and the result is that one actually gets $1/N$ times the
right moving part of the heterotic $SO(32)$ partition function with $q$
replaced by $q^{\frac{1}{N}}$. Including also other elements $h \neq 1$,
the final result is: 
\be 
\frac{1}{\tau_2^4} \frac{1}{N} \sum_{s=0}^{N-1} Z(\omega^s q^{\frac{1}{N}}),  
\label{longstring} 
\ee
where $Z(q)$ is
the right moving part of the $SO(32)$ heterotic partition function:  
\be 
Z(q) = \frac{1}{\eta(q)^{24}} \sum_{P \in \Gamma_{16}} q^{\frac{1}{2} P^2}, 
\ee
with $\Gamma_{16}$ is the ${\mathrm spin}(32)/Z_2$ lattice.

For the sector with $[g] = ((L)^M, \epsilon)$, we can again repeat the
above steps. Recall that in this case there are only two possible
$\epsilon$ that give non zero contribution to the elliptic genus. These
are given by $\epsilon= \epsilon_1.\epsilon_2\dots\epsilon_M$, with all
$\epsilon_i$ either an even or odd element of $Z_2^L$. As described above,
the order of the centralizer $C_g$ is $M! L^M 2^M$, while the number of
elements $h\in C_g$ that give rise to non-zero trace is $(M-1)! L^M 2^M$,
and therefore these are the relevant elements for the computation of
elliptic genus.  However, not all the $h$'s of this form give different
traces. Indeed, if $h$ and $h'$ are in the same conjucacy class in $C_g$,
they will give the same trace. It is easy to verify that the number of
elements in the centralizer ${\hat{C}}_h$ in $C_g$, for a relevant $h$, is
$2ML=2N$. As a result, the number of elements in the conjugacy class of
such $h$ in $C_g$ is $|C_g|/|{\hat{C}}_h| = (M-1)!L^{M-1} 2^{M-1}$. The
distinct conjugacy classes, that give non-zero traces, can be labelled by
$[h_i^{\pm}]$ for $i=1,\dots,L$, and the superscript $\pm$ refers to the
GSO projection on the $\chi$'s.  Each of these classes appear with a
prefactor, which is given by the number of elements in the class divided
by the order of $C_g$, and is equal to $1/(2N)$. The factor $1/2$,
together with the GSO prjection implied in $\sum_{\pm} tr [h_i^{\pm}]$,
for each $i$, gives either the scalar or spinor conjugacy classes of
${\mathrm spin}(32)/{\bf Z}_2$. The scalar and the spinor classes of
${\mathrm spin}(32)/{\bf Z}_2$ appear for the two choices of $\epsilon$ in
$[g]$. For each distinct $i$ in $\sum_i tr [h_i^{\pm}]$, one gets a
different trace, and the elliptic genus, after some tedious algebra, turns
out to be: 
\be 
\frac{1}{\tau_2^4}\frac{M^4}{N} \sum_{s=0}^{L-1} Z(e^{2\pi i \frac{s}{L}}
q^{\frac{M}{L}}).
\label{intstring} 
\ee
Here, the prefactor $M^4$ appears due to the zero modes of $8(M-1)$ $S$'s
in the $[g]$ twisted sector (excluding the center of mass $S$). Indeed,
these contribute to the trace as $\sqrt{\prod_{j=1}^{M-1}(1-e^{2\pi i
j/M})^8} = M^4$. Note that when $M$ and $L$ are not coprime, the different
terms in the sum above are not all related by modular transformations of
the type $\tau \rightarrow \tau + s$. It is, however, clear that each
term, which survives the projection, comes with integer multiplicity, as
it should be.

So far we had not turned on the Wilson lines $B$ in the Cartan subalgebra
of $SO(32)$. The presence of these Wilson lines twists the boundary
conditions for $\chi$'s. The determinants for the $\chi$'s with these
additional twists can be calculated in the standard way and, after
including the zero point shifts, one finds that, for the longest string
sector, the result is: 
\be 
Z(1,N)= \frac{1}{\tau_2^4} \frac{1}{N} \sum_{s=0}^{N-1} \frac{1}{\eta(\omega^s
q^{\frac{1}{N}})^{24}} \sum_{P\in \Gamma_{16}}\omega^{s\frac{P^2}{2}}
q^{\frac{1}{2N} (P+ NB)^2} ,
\label{1N} 
\ee 
and for the intermediate strings: 
\be 
Z(L,M)=\frac{1}{\tau_2^4} \frac{M^4}{N} \sum_{s=0}^{L-1}
\frac{1}{\eta(e^{2\pi i \frac{s}{L}} q^{\frac{M}{L}})^{24}} \sum_{P\in
\Gamma_{16}}e^{2\pi i\frac{sP^2}{2L}} q^{\frac{M}{2L} (P+ LB)^2}. 
\ee
Note that in the presence of Wilson lines, the different terms in the sum
over $s$ are not obtainable from the $s=0$ term by modular transformations
$\tau \rightarrow \tau + s$. This is so, because the Wilson lines, which
are turned on only in the $\sigma$ direction, break the symmetry between
the $\sigma$ and $t$ directions. 

\section{Longest string versus intermediate or short strings}

Even though the computation of the elliptic genus received contributions
from both the longest string sector and the intermediate or short string
sectors, it is only the longest string sector that corresponds to the
threshold bound states of $N$ D-strings. Consider for example the shortest
string sector. It receives contributions only from the sector $\tr h
(-1)^F$ where $[h]=(N)$. In this sector only the states having zero
relative transverse momenta survive. In position space, therefore, the
wavefunction of each of the $N$ strings is constant along the relative
separations. As a result such states are not normalizble. The same
argument also applies to the intermediate strings. In this case there are
$M$ groups of strings of length $L$ each and the wavefunction is constant
as a function of the relative separations between these $M$ groups. This
state therefore represents a state of $M$ strings, each of which is a
threshold bound state of $L$ strings.  The analogue of a single particle
state appears only in the longest string sector with $M=1$ and $L=N$. This
interpretation is also clear intuitively from the orbifold conformal field
theory description, since the twisted states in this sector correspond to
wavefunctions which are localized at the fixed point.

To see this more clearly, we can compactify one of the transverse
directions, say $X_8$, on a circle of radius $r$ and give the system a
total momentum $1/r$ along this direction. Note that this is the minimal
unit of quantized momentum. We will now show that only the longest string
sector can carry this momentum. 

The zero modes for $X_8^i$ ($i=1,\dots,N$) for a general twisted sector of
relevance labelled by $(L,M)$ reads, after suitable ordering of indices, as:
\be
X_8^i = a^i + \frac{kr}{M} \frac{2\pi t}{T} +\frac {\ell r}{L}
\frac{\sigma}{R_I}, \ee where $a^i$ satisfy \begin{eqnarray} a^{i+L} &=&
a^i + \frac{2\pi kr}{M} , \nonumber \\ 
a^{j+1} &=& a^{j} + \frac{2 \pi
\ell r}{L}, ~~~~~~ j=1,\dots,L-1, 
\end{eqnarray} 
and $k$ and $\ell$ are
arbitrary integers. Note that the integers $k$ and $\ell$ are independent
of $i$ because only the center of mass $X_8$ is a zero mode under the
combined actions of the twists along the $t$ and $\sigma$ directions.
$\ell$ here denotes the winding number along $X_8$ direction. These zero
modes contribute to the action as 
\be 
\Delta S =
\frac{\pi}{\alpha'_{\mathrm eff}}[ \frac{N}{\tau_2} (\frac{kr}{M})^2 +
N\tau_2 (\frac{\ell r}{L})^2 ].  
\ee 
The total momentum is $\frac{1}{2\pi
i \alpha'_{\mathrm eff}}\int d\sigma \sum_i \partial_t X_8^i =
\frac{Nkr}{Mi\tau_2\alpha'_{\mathrm eff}}$. We can now perform a Poisson
resummation in order to go to the Hamiltonian formulation, with the result
that the total momentum $p$ along the $X_8$ direction is:  
\be 
p = \frac{Mk}{r} , 
\ee 
with $k$ some integer and the partition function is 
\be
\sum_{k,\ell} q^{\frac{p_L^2}{4\alpha'_{\mathrm eff}N}}
\bar{q}^{\frac{p_R^2}{4\alpha'_{\mathrm eff}N}}\sqrt{\tau_2} Z(L,M) , 
\ee
where $p_L= M(\frac{\alpha'_{\mathrm eff}k}{r} + \ell r)$ and $p_R =
M(\frac{\alpha'_{\mathrm eff}k}{r}-\ell r)$.  This shows that the smallest
unit of momentum, $p=1/r$, gets contributions only from the $M=1$ (i.e.
the longest string). From the above partition function we see that this
state carries an extra energy given by $\alpha'_{\mathrm eff}p^2/2NR_I$
which, as we shall see below, is exactly what is expected from the
heterotic string side. When $p=M/r$, the $(L,M)$ sector also contributes
to the elliptic genus.  However, this is consistent with the
interpretation that it corresponds to $M$ groups of strings, each carrying
a momentum $1/r$. 

\section{Comparison with the heterotic spectrum}

Let us now compare the above results for the bound states of $N$ type I
D-strings wrapped around the $X^9$ circle with the heterotic spectrum in 9
dimensions. Clearly the relevant states on the heterotic side are the ones
carrying a winding number $N$. The remaining quantum numbers are the
$U(1)^{16}$ charges of the Cartan subalgebra of $SO(32)$ and the
Kaluza-Klein momentum along the $X^9$ direction. The $U(1)^{16}$ charges,
on the type I side, can be read off from the $\Gamma_{16}$ lattice charges
that appear in the partition function $Z(1,N)$. The Kaluza-Klein momentum
$p_9$, on the other hand, is the longitudinal momentum $P_{\sigma}$ of the
D-string system along the $\sigma$ direction. Given the fact that
$P_{\sigma}$ is the difference of the left and right Virasoro generators
$L_0-{\bar{L}}_0$, its charge is just given by the coefficient of $ 2\pi i
R_I\tau_1$ in the partition function. From the expression (\ref{1N}) for
$Z(1,N)$ and taking into account the projection implied by the sum over
$s$, we conclude that
\begin{eqnarray}
P_{\sigma} &=& \frac{1}{R_I}( k + B.P + \frac{1}{2} B^2 N ),  \nonumber \\
k &=& \frac{1}{N} (\frac{1}{2}P^2 + N_R -1) \in {\mathbf Z},
\end{eqnarray}
where $(N_R-1)/N$ appears from the expansion of
$\eta(q^{\frac{1}{N}})^{-24}$. Note that the multiplicity of these states
is the same as the coefficient of $q^{N_R-1}$ in the expansion of
$\eta(q)^{-24}$. The value of $k$ is bounded below by $(-1)$ for $N=1$ and
by $0$ for $N > 1$, while the value of $P_{\sigma}$ is bounded by
$-1/NR_I$.  These two equations are exactly the ones appearing in
(\ref{kkmomI}) for the Kaluza-Klein momentum and the level matching
condition (\ref{level}) for the BPS states. It is also clear that the two
multiplicities match, as both are given by the coefficient of $q^{N_R-1}$
in the expansion of $\eta(q)^{-24}$. 

Furthermore, the mass of the bound state is the original mass of $N$
D-strings wrapped around the circle, plus the energy carried by the
excitation, which is given by the coefficient of $T = R_I\tau_2$ in
$Z(1,N)$. Since the partition function depends only on $q$ the latter is
equal to $P_{\sigma}$. Thus the total energy is $\frac{NR_I}{\alpha'
\lambda_I} + P_{\sigma}$. This is exactly the mass given in (\ref{massI})
predicted by the duality upto a sign. As mentioned earlier $P_{\sigma}
\geq -1/NR_I$. Therefore, for $R_I^2 > \alpha' \lambda_I/N^2$ the quantity
$\frac{NR_I}{\alpha' \lambda_I} + P_{\sigma}$ is positive definite and
hence it coincides with the absolute value appearing in (\ref{massI}).
However, for $R_I^2 < \alpha' \lambda_I/N^2$ this quantity is negative,
for a suitable choice of the Wilson line $B$, and the result would not
make sense. But this is exactly the region in which the type I
perturbation theory breaks down, as argued in \cite{pw}. 

Finally, we consider the situation discussed in the last section where a
transverse direction is compactified on a circle of radius $r_I$ and the
system carries a momentum $k/r_I$. This does not alter the level matching
condition and therefore the multiplicity of the state.  Recalling that
$\alpha'_{\rm eff}=\alpha' \lambda_I$, we find that the extra energy is
$k^2 \alpha' \lambda_I/2NR_I r_I^2$. On the heterotic side the mass for a
state with winding number $N$ along the $X_9$ direction and carrying
momentum $k/r_H$ along the $X_8$ direction is given by 
\be
\frac{1}{\alpha'} \sqrt{N^2 R_H^2 + (\frac{k\alpha'}{r_H})^2} .  
\ee 
By using the duality relations (\ref{drel}) and expanding the square root
to the leading order in $\lambda_I$ we find that the extra mass is exactly
$k^2 \alpha' \lambda_I/2NR_I r_I^2$, in agreement with the prediction of
duality.

To conclude, we have analysed in detail the orbifold conformal theory
arising in the infra-red limit of the $O(N)$ gauge theory that describes
the low energy modes of a system of $N$ type I D-strings. We argued that
the longest string sector of the orbifold theory describes the bound
states of D-strings. An additional support for this identification also
comes by compactifying one of the transverse directions. We have shown
that this identification gives masses and multiplicities of the bound
states in agreement with the predictions of heterotic-type I duality in 9
dimensions, for all the BPS charges in $\Gamma_{(1,17)}$. In particular
the Kaluza-Klein momentum of the heterotic theory is mapped to the
longitudinal momentum of the D-strings.  \vskip .3in \noindent {\bf
Acknowledgements}

This research has been supported in part by EEC under the TMR contracts
ERBFMRX-CT96-0090 and FMRX-CT96-0012.

\rnc{\Large}{\normalsize}

\end{document}